# HIGH POWER DENSITY TEST OF PXIE MEBT ABSORBER PROTOTYPE

A. Shemyakin [#], C. Baffes, Fermilab[*], Batavia, IL 60510, USA

## Abstract

One of the goals of the PXIE program at Fermilab [1] is to demonstrate the capability to form an arbitrary bunch pattern from an initially CW 162.5 MHz H- bunch train coming out of an RFQ. The bunch-by-bunch selection will take place in the 2.1 MeV Medium Energy Beam Transport (MEBT) [2] by directing the undesired bunches onto an absorber that needs to withstand a beam power of up to 21 kW, focused onto a spot with a ~2 mm rms radius. Two prototypes of the absorber were manufactured from molybdenum alloy TZM and tested with a 28 keV DC electron beam up to the peak surface power density required for PXIE, 17W/mm$^2$. Temperatures and flow parameters were measured and compared to analysis. This paper describes the absorber prototypes and key testing results.

## PXIE ABSORBER CONCEPT AND PROTOTYPING

The PXIE MEBT chopper concept [2] assumes that broadband kickers direct the undesired bunches onto an absorber by shifting them from the "unperturbed' trajectory by 6 rms beam sizes (Fig.1A). The absorber concept [3] features a grazing angle of incidence (29mrad) that decreases the peak absorbed surface power density to 17 W/mm$^2$; longitudinal segmentation to relieve thermally-induced stresses (Fig.1B); and molybdenum alloy TZM as the choice for the absorbing surface. The power density estimate assumes that 25% of the beam power is reflected.

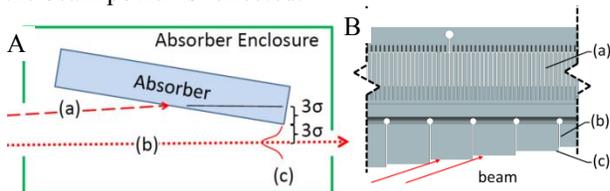

Figure 1: Concept the MEBT absorber. A- separation scheme with (a) chopped beam, (b) passed beam, (c) 6σ shift between the chopped and passed beams. B- absorber concept features: (a) cooling microchannels, (b) stress relive slits, (c) shadowing steps (magnitude exaggerated).

In this design concept, longitudinal heat transfer is interrupted by stress relief slits, and heat flows primarily transversely. Therefore, the risks related to the high surface power density can be addressed with a short prototype. Two such prototypes were manufactured and tested with an electron beam at a dedicated test stand.

## TEST STAND

At the stand (Fig.2), the prototypes were heated by an 27.5keV, typically 0.19A electron beam. The beam size was adjusted by solenoids, and it was positioned by dipole correctors.

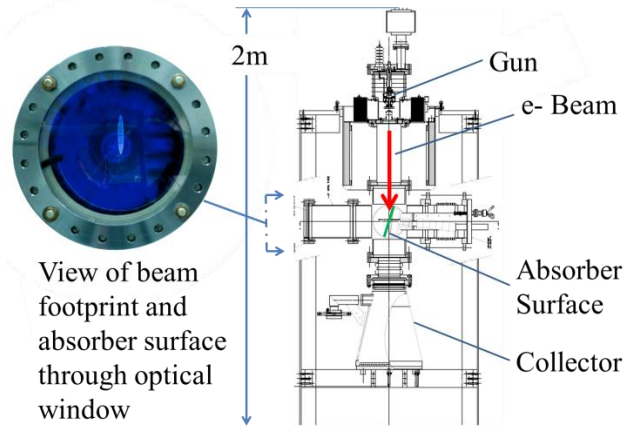

Figure 2: The test stand and OTR beam image

The coolant temperature and flow were measured while the beam was directed either to the prototype or into an effective beam collector. Comparison of caloriemetry in these cases showed that a significant portion of the beam power (~55%) was reflected or carried away in secondary electrons, so the power absorbed in the prototype was ~2.4kW.

The images of the prototype surface were recorded by a TV camera through a quartz viewport. The beam interaction with the surface produces a visible light with intensity proportional to the beam current, which was interpreted as Optical Transition Radiation (OTR). At elevated surface temperatures, the images were dominated by thermal radiation. Recording images with optical filters of different central wavelength allowed separating the broad band OTR signal from typically longer wavelength thermal radiation. While the former provided information about the beam profile, the latter helps to estimate the surface temperature (see details in Ref. [4]). Neutral filters (1% and 10%) were used to increase the dynamic range.

## PROTOTYPE I

The absorbing surface of the first prototype (Fig.3) was manufactured from a monolithic TZM. Coolant flows transversely through the absorber body in 300μm wide cooling channels fabricated by EDM. A more comprehensive description of the Prototype I is given in Ref. [4], [5].

Over several months of testing, the beam size was decreased until the peak power density expected for PXIE of 17W/mm2 was achieved over one "fin", a fin being an area of the surface bordered by stress relief slits. In this condition, peak surface temperatures of ~1300K were reconstructed from optical measurement and analysis. After exposure to this thermal condition, the absorber survived and did not exhibit any symptoms of damage. This is the primary result of this testing program.

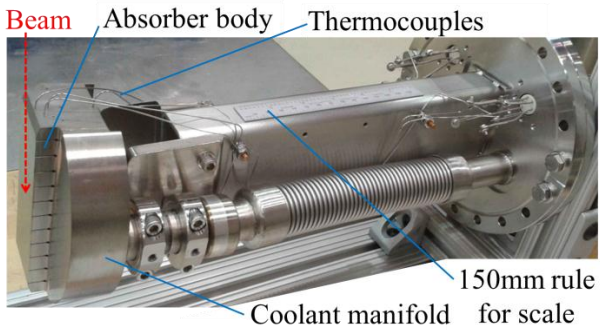

Figure 3: Absorber Prototype I.

For studied focusing conditions, visible radiation and thermocouple temperature profiles were captured for comparison with analysis. Given the OTR-derived beam profile (Fig.4) and total energy deposition estimated from calorimetry, finite element analysis was performed in ANSYS [5] to predict TZM temperatures, both at the surface and at the discrete thermocouple locations within the absorber body.

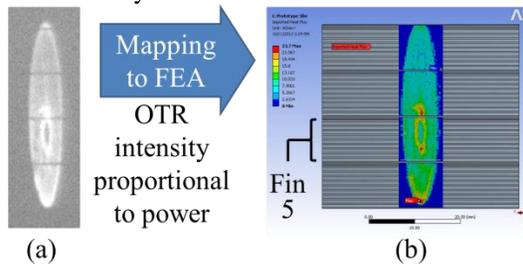

Figure 4: Reconstruction of energy deposition at intermediate focusing condition. (a) – Blue filtered OTR image (b) – Analyzed beam profile. Power density on Fin 5 10 W/mm$^2$ average, 25 W/mm$^2$ peak.

Predicted surface temperatures were compared to temperature estimates calculated from thermal radiation (Fig.5).

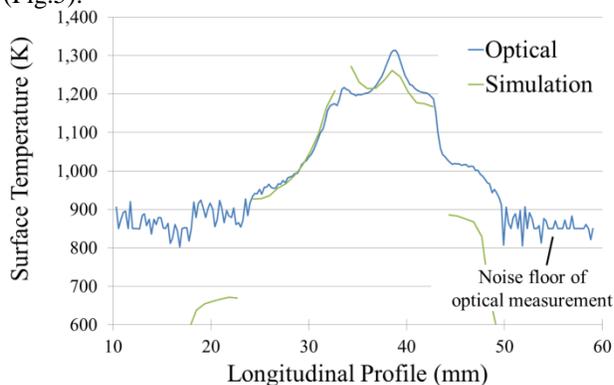

Figure 5: Comparison of FEA simulation and optical measurement of surface temperature along a linear profile at maximum (for Prototype I) power density.

As described in [4], fitting parameters used in the radiation calculations were adjusted to optimize fit with FEA simulation, so any agreement better than the inherent uncertainty of the optical measurement (±150K) is, to some extent, enforced.

Temperatures within the absorber body were monitored by an array of thermocouples. Simulations agreed well with thermocouple measurements, and tended to under-predict the observed temperature rise by 3-10%.

Thermal performance of the cooling scheme was tested over a range of flow rates with velocity through micorchannels varying between 0.6 and 3.5m/s. No indications of boiling were observed.

## PROTOTYPE II

While Prototype I was tested successfully, its fabrication and testing revealed some deficiencies in the design. Better management of reflected energy is desirable. Having water flow through the inherently brittle TZM required a complex design and fabrication and created the possibility of water-to-vacuum failure mode. Given the good thermal performance of the Prototype I, the authors were emboldened to sacrifice some thermal performance in favor of a simpler design. The design of Prototype II (Fig. 6) relies on thermal contact between the TZM absorbing surface and a cooling block made of aluminum. The absorbing surface is assembled from 6 separate pieces (fins).

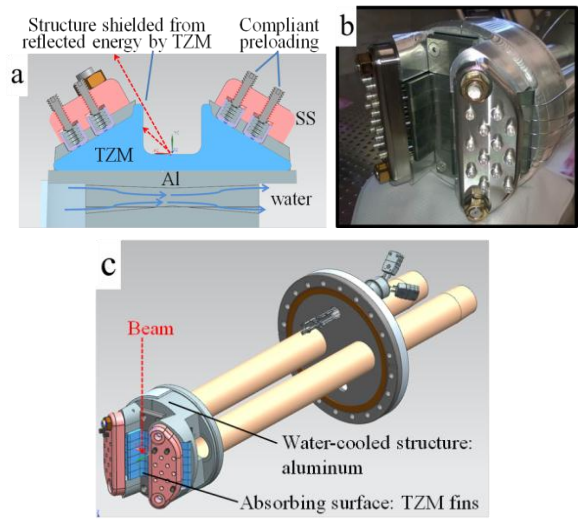

Figure 6: Prototype II. a- cross section. Beam is into page. b- photo of the assembled prototype. c- 3D model. 6 fins are shown in blue.

At the preloaded joint between TZM and aluminum, the thermal contact is enhanced by a compliant graphite interface foil. The TZM absorbing surface extends to include side "walls." These walls reabsorb some of the particles reflected from the absorber surface, and in the H- PXIE application limit the areas of the absorber enclosure where cooled and blistering- resistant secondary absorbing surface must be provided. Profile of water channels is identical to Prototype I's.

The Prototype II was tested with the same technique as described in the previous section. The maximum tested power density averaged over entire beam footprint was 23 W/mm$^2$. The surface temperature was estimated from optical measurements using coefficients found in the Prototype I fitting. The maximum temperature was 1500 – 1700K (Fig. 7), in a reasonable agreement with simulations. Note that at comparable conditions (as in Fig.5), the surface temperature was by ~150K higher than during Prototype I testing.

The thermocouples in Prototype II were positioned in channels between the fins (as opposed to being placed in longitudinal channels in Prototype I) to better mimic the possible arrangement in the full-length absorber and simplify the assembling. It was found that in a case of a significant longitudinal temperature gradient due to inhomogeneity of the beam current density, thermocouples contact with neighboring fins varied with different beam positions and sizes, which added complexity in interpretation of the results. Typically simulations predicted the thermocouple temperatures higher by 10-20% than those found in measurements.

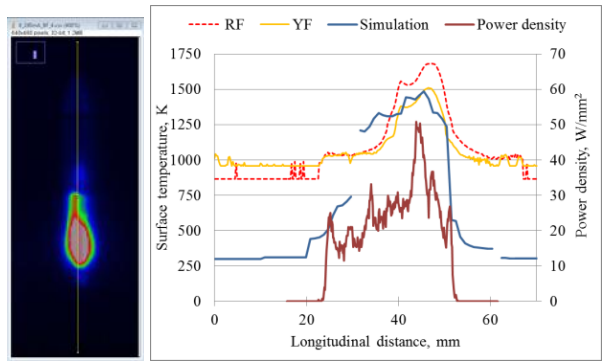

Figure 7: Left: image of the beam footprint with maximum power density recorded with a blue filter. False colors in ImageJ [7]. The yellow line shows where the profile was analyzed. Right: the temperature profiles reconstructed from images recorded with red and yellow (plus neutral) filters as well as the power density distribution.

Comparison of simulations with measurements allowed estimating the most important parameter of the Prototype II test, the thermal contact through graphite at operational conditions, within a factor of 2. The best fitting value is $2 \cdot 10^4$ W/(m$^2$K). The surface temperature profiles along the beam footprint were similar for various positions of the beam, which implies that the thermal contact was the same for different fins within the measurement uncertainty.

Presently the Prototype II is under thermo-cycle testing. If it is successful, the further tests will be conducted in the PXIE beam line. The Prototype II is expected to be exposed to a 2.1 MeV H- beam with the average power up to ~5 kW in CY2015. Note that the peak power density expected at PXIE is noticeably lower than has been demonstrated at the test stand (Fig. 8), and the main issues to address will be resistance to H- induced blistering and operational peculiarities (e.g. tuning in absence of OTR imaging).

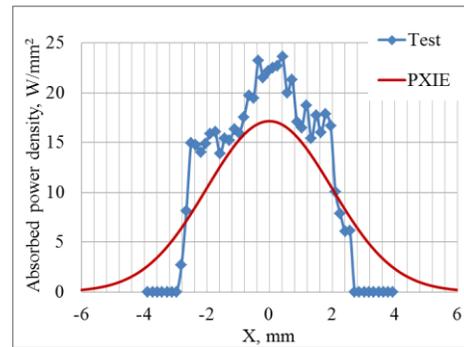

Figure 8: Comparison of transverse absorbed power density profiles expected for the PXIE absorber and used in thermocycling tests.

## CONCLUSION

Two absorber prototypes were successfully tested to the power density exceeding the specification for the PXIE absorber. The concept of the second version, where the thermal performance was sacrificed for simplicity and robustness, is the leading candidate for the full-length absorber design.

## ACKNOWLEDGMENTS

The authors wish to acknowledge the efforts of K. Carlson, C. Exline, B. Hanna, A. Mitskovets, L. Prost, K. Reader, R. Thurman-Keup and J. Walton in assembling, commissioning, and operating the test stand.